\documentclass[floatfix,amsmath,amssymb,amsfonts,bbm,pra,twocolumn,letterpaper]{revtex4}
\usepackage[latin1]{inputenc}
\usepackage{amsmath,hyperref}
\usepackage{amsfonts}
\usepackage{bbm}
\usepackage{verbatim}

\newcommand{\ket}[1]{|#1\rangle}
\newcommand{\bra}[1]{\langle #1|}
\newcommand{\bracket}[2]{\langle #1|#2\rangle}
\newcommand{\ketbra}[1]{|#1\rangle\langle #1|}

\newcommand{\minus}{\!-\!}

\begin{document}

\title{Generalized decoding, effective channels, and simplified security proofs in quantum key distribution}

\author{Joseph M. Renes$^{1,2}$ and Markus Grassl$^1$}
\affiliation{$^1$IAKS Prof.~Beth, Arbeitsgruppe Quantum Computing, Universit\"at Karlsruhe, Am Fasanengarten 5, D-76131 Karlsruhe, Germany\\
$^2$Quantum Information Theory Group, Institut f\"ur Theoretische Physik I and 
Max-Planck-Forschungsgruppe, Institut f\"ur Optik, Information und Photonik,\\ 
Universit\"at Erlangen-N\"urnberg, Staudtstrasse~7, D-91058 Erlangen, Germany}

\begin{abstract}
Prepare and measure quantum key distribution protocols can be decomposed
into two basic steps: \emph{delivery} of the signals over
a quantum channel and \emph{distillation} of a secret key from the signal 
and measurement records by classical processing and 
public communication. Here we formalize the distillation process for a 
general protocol in a purely quantum-mechanical framework and 
demonstrate that it can be viewed as creating an ``effective'' quantum 
channel between the legitimate users Alice and Bob. The process of secret 
key generation can then be viewed as entanglement distribution using this channel, 
which enables application of entanglement-based security proofs to 
essentially any prepare and measure protocol. 
To ensure secrecy of the key, Alice and Bob must be able to estimate 
the channel noise from errors in the key, and 
we further show how symmetries of the distillation process simplify this task. 
Applying this method, we prove the security of several 
key distribution protocols based on equiangular spherical codes. 
\end{abstract}

\maketitle

\section{Introduction}

Quantum key distribution (QKD) is currently the most successful 
theoretical and practical application of quantum information theory to solving a 
real-world problem that classical information theory cannot: 
secure expansion of previously held keys between two separated parties
using public channels.
In its simplest form, it only requires that one party, Alice, prepare
and send individual quantum systems to the other, Bob, who immediately
measures them. No collective storage or manipulation of the quantum
systems is required, making it a very humble foray into the quantum
world.  After the quantum communication phase is complete, Alice and
Bob have classical strings corresponding to the signal and measurement
records, respectively. With the aid of a public classical channel and their
previously held keys, they can then collaborate to distill the new
(longer) secret key from these strings.

Ever more sophisticated methods of proving the unconditional 
security of such protocols have recently been developed. 
In particular, strong links have been forged between the security 
of a given protocol and the ability of a suitable quantum version
to implement entanglement distillation. Building on work by Lo and 
Chau~\cite{lc99}, Shor and Preskill~\cite{sp00} demonstrated that
the classical distillation steps 
of traditional prepare and measure schemes could be seen as a version
of entanglement distillation by using Calderbank-Shor-Steane (CSS) quantum 
error-correcting codes~\cite{css}. They illustrated this technique 
by application to the prototypical Bennett-Brassard 1984 (BB84) protocol~\cite{bb84}, 
and the analysis of the structurally similar six-state protocol~\cite{bruss98} 
followed soon thereafter~\cite{l01}. 
By viewing the measurement as a local filtering operation~\cite{Filtering}, 
the Bennett 1992 (B92)~\cite{b92,tki03} and ``trine'' Phoenix-Barnett-Chefles 2000/Renes 2004 (PBC00/R04)~\cite{pbc00,renes04,btblr05} 
protocols were tackled by essentially orthogonalizing some of the measurement outcomes 
in order to prepare them for the CSS-based error correction. 

The main obstacle to formulating such an unconditional security proof for general 
protocols is the conceptual difficulty of reconciling the framework of entanglement 
distillation with the requirements proscribed by the protocol. 
How to perform entanglement distillation is clear enough; the trick here is 
to apply it to the correct quantum state such that the entire process properly
mimics the actual prepare and measure protocol.

Put differently, the problem lies in providing a quantum description of the so-called ``sifting'' 
operation in which the signal and measurement records collected during the 
quantum communication phase are transformed into a raw key. 
The name comes from the BB84 protocol, where Alice and Bob 
keep only those signals and measurements
for which the associated bases used in preparation and measurement match, 
thus sifting the ``good'' bits from the ``bad.''
The use of local filtering, as in the analysis of the B92 and trine protocols, is one 
possible quantum description of the sifting, or, more generally, \emph{decoding}, step.
However, it implicitly assumes that the distillation process requires only one-way 
communication from Alice and Bob.

In this paper, we develop a general quantum-mechanical formulation of the decoding step
applicable to a broad class of key distribution protocols. This immediately leads to
a general framework for unconditional security based on entanglement distillation,
which we illustrate by proving the unconditional security 
of a several equiangular spherical code protocols.
Formalizing the decoding step in this manner offers insight into 
the mechanism underlying key distribution protocols.
From this vantage point we see that the decoding step performs two critical tasks. 
First, the physical quantum channel and the decoding process merge into an ``effective'' or logical
quantum channel connecting Alice and Bob. This channel describes how the 
physical signal system is transformed into a logical key system. 
Second, noise in the physical quantum 
channel caused by an eavesdropper, Eve, will be mapped to noise in the effective channel.
Only the latter is relevant, as it is related to what Eve might know
about the key. Moreover, this noise can be easily 
estimated from the error rate observed in the decoded key string and can then be 
used to ensure the security of the protocol. 

The remainder of the paper is organized as follows. 
Section~\ref{sec:pandm} lays out the 
details of the prepare and measure schemes under 
consideration. Section~\ref{sec:qf} then presents a 
fully quantum-mechanical formulation of the decoding phase.
Using this, Sec.~\ref{sec:ec} details how the 
decoding step creates an effective quantum channel between Alice and Bob 
within the postselected state space, with simplified 
noise patterns relative to the actual physical channel. 
The resulting formalism then enables us to easily treat the question of 
security for more-complicated and higher-dimensional protocols.  
Section~\ref{sec:secproofs} is specifically devoted 
to the security of protocols using equiangular spherical codes in two and three dimensions. 
Finally, Sec.~\ref{sec:dc} concludes with a discussion 
of further applications of this work and open problems.

\section{Prepare and Measure Protocols}
\label{sec:pandm}
In a generic prepare and measure quantum key distribution protocol, 
two separated parties, Alice and Bob, wish to make use of an insecure quantum
channel and a classical public broadcast channel in order to establish a shared, secret string. 
They already share a short key with which they can authenticate messages 
from each other sent on the classical channel. The goal is to expand this short 
key into a longer version, suitable for encrypting a sizable amount of data. 
Roughly, their strategy is to use the quantum channel to distribute
quantum states, which can then be translated into a (classical) raw key. 
From this substrate the final key can be distilled with the aid of 
communication over the classical channel. By using quantum states, they will be able to 
quantify the effect of Eve's interference so that the appropriate 
countermeasures may be taken during the distillation step---e.g.,~privacy amplification. 
In the worst case, they can abort the protocol if they find that Eve's spying 
on the channel is so severe that no secret key can be created.

These sorts of key distribution protocols can be decomposed into two phases: 
a delivery phase using the quantum channel and a
distillation phase using the classical channel. 
Alice sends signals to Bob over the quantum channel in the delivery phase, who
immediately measures them---hence
the term ``prepare and measure.'' The signals are drawn from the ensemble of signal letters
$\mathcal{S}=\{\ket{\xi_j}\in\mathbb{C}^d\}_{j=1}^{n}$, 
where the prior probability for each signal is encoded in its squared norm: $\pi_j=\bracket{\xi_j}{\xi_j}$.
Bob's measurement is described by a positive-operator-valued measure (POVM) $\mathcal{M}=\{\ket{\eta_k}\in\mathbb{C}^d\}_{k=1}^{m}$ such that
$\sum_k\ketbra{\eta_k}=\mathbbm{1}$. Without loss of generality both 
$\mathcal{S}$ and $\mathcal{M}$ are ensembles of pure states since
ensembles of mixed states could be further decomposed into them. 

A signal ensemble is termed \emph{oblivious} when 
$\sum_j \ketbra{\xi_j}=\mathbbm{1}/d$, meaning that a random
signal on the quantum channel is completely unbiased. 
In contrast, general ensembles are biased, a property Eve may be able to exploit.
Here we will focus on oblivious ensembles with uniform 
prior probabilities; the obliviousness will play a small but important role in the 
next section.

Given a noiseless quantum channel, the joint probability for Alice to send 
the $j$th signal and Bob to obtain the $k$th outcome is given by the simple rule
\begin{equation}
\label{eq:prob}
p_{jk}=|\bracket{\eta_k}{\xi_j}|^2.
\end{equation}
Every round yields Alice and Bob one letter each; repeating the 
protocol generates strings which are samples from this joint distribution. 
These strings are the output of phase 1. 

The task of phase 2 is to distill these strings into a shared, secret key. 
This process can be represented by a pair of functions, one each for Alice and
Bob, which map the signal and measurement strings to key strings. 
An ordinary protocol will consist of several rounds of mappings, and in each round
the purpose of the classical communication is  to coordinate the application of the
associated functions.
The term ``decoding'' refers to the initial rounds of the distillation process, specifically
those required to produce a secret key given a noiseless channel.
Additional distillation steps are required for noisy channels---namely, information
reconciliation to correct mismatched key letters and privacy amplification
to ensure secrecy of the resulting key. 

The set of distillation functions is quite large and the choices of protocols myriad. 
For concreteness, we shall focus on maps which attempt to distill one key 
letter from each signal-measurement pair by use of one-to-one functions.
Note that the distillation procedure may, and often does, fail for particular inputs. After presenting 
and examining this formalism, we will describe how
to make generalizations for more complicated strategies. 

Here it is convenient to describe the distillation functions
via their inverses. Suppose that each of Alice's and Bob's maps
results in a letter drawn from the set $\{0,\dots, r\minus 1\}$. Naturally, $r\leq\min(n,m)$. 
The action of one of Alice's maps can be succinctly captured by the $r$-tuple 
$\big(\sigma(0),\dots,\sigma(r\minus 1)\big)$, where $\sigma(x)$ is the
input signal which led to the key letter $x$. For example, if Alice draws
signals from the set $\{a,b,c,d,e\}$ and a decoding function maps $b$ to 0 and
$d$ to 1, the corresponding tuple is simply $(b,d)$. The $r$-tuple thus records which 
inputs become which key letters. Note that in this convention the distillation map is 
$\sigma^{-1}$ and all inputs not appearing 
in the $r$-tuple are discarded. By a slight abuse of notation 
we denote this with the output symbol $\square$; for instance, in the
previous example $\sigma^{-1}(a)=\square$. 
Altogether, we shall assume that  Alice has $n_a$ (inverse) 
distillation functions $\sigma_s$, while Bob 
has $n_b$ functions named $\tau_t$. 

The set $T$ of allowable function pairs $(s,t)$ fully describes each distillation step. 
Alice and Bob use the classical channel to coordinate their actions and determine
if the applied function pair yields a key 
letter---legitimate function pairs may still fail to produce a key 
letter for a given input. Again there are several options in how to 
accomplish this in practice; here, we adopt a particular communication 
scheme to perform allowable decodings without 
making any claim to its generality. One of the parties---say, Alice---initiates 
the procedure by randomly choosing a function $s$ compatible 
with her signal---i.e., a function which does not map the signal to 
$\square$---and announcing this choice to Bob. 
He can then infer 
which of his functions ensures that $(s,t)\in T$ and then randomly apply one
of them. 

The BB84 protocol provides the simplest example of this framework. 
$\mathcal{M}$ and $\mathcal{S}$ 
both consist of (appropriately normalized) linear polarization states, 
either horizontal or vertical or inclined at $\pm 45^\circ$. Label these states 
$\{\textrm{---},|,\diagup,\diagdown\}$. Only those signals and measurements
belonging to the same basis are to be kept, 
so the possible sifting functions for both Alice and Bob are 
represented by the tuples $(\textrm{---},|), (|,\textrm{---}), (\diagup,\diagdown)$, and $(\diagdown,\diagup)$. 
The set $T$ just consists of the same function for each party. 
To perform the decoding, Alice applies either of the two applicable functions
to each signal and sends a record of her action to Bob. This
tells him which decoding function to use, and if applying it to his measurement result
does not produce the output $\square$, Bob keeps the output and tells Alice. 

Generalizations to more complicated schemes are now 
straightforward. Keeping to single-letter decoding, 
whole sets of signal or measurement letters can be mapped to 
different key letters simply by considering $r$-tuples whose entries are these sets. 
The modifications to the functions for block decoding 
are self-evident: block inputs and block outputs, keeping the reject output $\square$. 
To illustrate the latter, consider a parity-check advantage distillation step~\cite{maurer93}. 
Alice computes the parity of a particular pair of letters and transmits this to Bob. 
If this matches the parity of his corresponding pair, they keep the first letter; otherwise,
they discard both. In the present framework, this is described compactly by the
decoding tuples $(00,11)$ and $(01,10)$, where $T$ consists of the same tuple for each party.
 
As the distillation process is meant to create not only a correlated, but also secret string, 
we must consider the effect of announcing the distillation function publicly, which could
reveal information to Eve. Chosen properly, however, the decoding will leave Eve 
completely nescient of the key. Such is the case in the BB84 protocol, where the sifting information
tells Eve that the signal was one of two possibilities, instead of the four originally
possible. Due to the structure of $T$, this information is completely independent 
of the resulting key bit. 

If the public communication leaks no information about the key, then Eve's probability 
for the key must be uniform. As a sufficient (but not necessary) condition, we can require both that the 
signals be chosen with uniform probability and that, in the multiset formed by the 
union of all of Alice's $r$-tuples, each signal appears the same number $n_a$ times. 
To cover the cases in which the information flows from Bob to Alice, we will assume 
that the measurement outcomes each appear the same number $n_b$ times in his multiset. 
This includes essentially every proposed key distribution protocol and, like the choice of 
oblivious signal ensemble, will have some advantages in the next section.  

This limits Eve's source of information about the key to the quantum channel. The decoding 
process will turn channel noise into key errors, and the number of errors 
in the decoded key will be linked to the amount of information Eve could 
in principle obtain. By measuring the error rate, Alice and Bob can tailor the remaining 
distillation steps to suit their needs. For the prepare and measure schemes under consideration here, 
we assume that the further processing is independent of the specific decoding details. 
That is to say, after making the decoded key, Alice and Bob forget 
which key letters were the output of which decoding functions. As much is 
done in the BB84 protocol, for example; basis information is irrelevant after the sifting phase.
This is not a trivial step, 
since by retaining complete information, Alice and Bob could possibly find that key letters 
from certain decodings require different handling than others. 
However, it is not only vastly simpler to consider the average case, but also affords 
considerable simplification of the channel noise, as discussed
in Sec.~\ref{sec:ec}. 
 
\section{Quantum Formulation}
\label{sec:qf}
We now give a fully quantum-mechanical description for the prepare and measure protocol.
In doing so, we must retain the essential features of the protocol---namely, the type of physical 
system actually sent and the distribution of signals and measurements, given by Eq.~(\ref{eq:prob}). 
Let Alice begin with the state
\begin{equation}
\label{eqn:phient}
\ket{\Phi}=\sqrt{d}\sum_j\ket{\xi^*_j}_A\ket{\xi_j}_B\in\mathcal{H}_{\rm phys}\otimes\mathcal{H}_{\rm phys},
\end{equation}
where $\ket{\xi^*_j}$ is simply the complex conjugate of $\ket{\xi_j}$ in the standard basis. 
The vector space to which $\ket{\Phi}$ belongs is explicitly given as it will prove useful to keep 
the various spaces clearly distinct. Here ``phys'' stands for ``physical,'' denoting that this
is the space which describes the actual physical signal sent. One may verify that $\ket{\Phi}$ is 
properly normalized by using the fact that $\bracket{\xi^*_j}{\xi^*_k}=\bracket{\xi_k}{\xi_j}$.

Since the signal ensemble is oblivious, Alice can prepare one of the signals $\ket{\xi_j}$ in
subsystem $B$ by measuring her half with the POVM
$\{d\ketbra{\xi_j^*}\}$.
Moreover, computing the expansion coefficients in the standard basis, we find
$\bracket{jk}{\Phi}=\delta_{jk}/\sqrt{d}$,
meaning that $\ket{\Phi}$ is the canonical maximally entangled state in $\mathbb{C}^d\otimes\mathbb{C}^d$. 

After distributing subsystem $B$ to Bob, they perform the following operations to their respective systems:
\begin{equation}
\label{eq:initstate}
P=\sqrt{d}\sum_j\ket{j}\bra{\xi_j^*},\quad M=\sum_k\ket{k}\bra{\eta_k},
\end{equation}
so that the state becomes
\begin{equation}
\label{eq:pmstate}
(P\otimes M)\ket{\Phi}=\sum_{jk}\ket{j}_A\ket{k}_B\bracket{\eta_k}{\xi_j}
\in\mathcal{H}_{\rm prep}\otimes\mathcal{H}_{\rm meas}.
\end{equation}
The partial isometries $P:\mathcal{H}_{\rm phys}\rightarrow\mathcal{H}_{\rm prep}$ and 
$M:\mathcal{H}_{\rm phys}\rightarrow\mathcal{H}_{\rm meas}$ realize the Neumark extensions 
of $\mathcal{S}$ and $\mathcal{M}$~\cite{nc00}. In other words, the POVM's on $\mathcal{H}_{\rm phys}$ are
promoted to projection measurements on $\mathcal{H}_{\rm prep}$ and $\mathcal{H}_{\rm meas}$,
all the while ensuring that the outcomes are still distributed according to Eq.~(\ref{eq:prob}).

Now for the crux of the whole enterprise. By promoting the POVM elements to projection
operators, each party's measurement can be easily restructured into two parts: a coarse-grained and
a fine-grained measurement. The coarse measurement is a projection onto a subspace spanned 
by many basis states, while the fine-grained measurement then locates the precise basis state in
the subspace. The crucial point is that the outcome of the coarse-grained measurement can 
be chosen to correspond to the distillation function. 

This is accomplished by employing the two operators 
$S_A:\mathcal{H}_{\rm prep}\rightarrow\mathcal{H}_{\rm a}\otimes\mathcal{H}_{\rm key}$ and
$S_B:\mathcal{H}_{\rm meas}\rightarrow\mathcal{H}_{\rm b}\otimes\mathcal{H}_{\rm key}$:
\begin{eqnarray}
\label{eq:siftops}
S_A\!&=&\!\sqrt{\frac{d}{n_a}}\sum_{sl}e^{i\theta(s,l)}\ket{s}\ket{l}\bra{\sigma_s(l)},\nonumber\\
S_B\!&=&\!\frac{1}{\sqrt{n_b}}\sum_{tm}e^{i\phi(t,m)}\ket{t}\ket{m}\bra{\tau_t(m)},
\end{eqnarray}
which relabel the $\mathcal{H}_{\rm prep}$ and $\mathcal{H}_{\rm meas}$ basis 
states in terms of two registers for the coarse- and fine-grained steps. 
The basis states of the vector spaces $\mathcal{H}_{\rm a}$ and $\mathcal{H}_{\rm b}$ 
label the decoding functions, while the vector spaces $\mathcal{H}_{\rm key}$ 
contain the decoded key. This is an equivalent representation of the 
state as long as the operators are partial isometries
(with $\mathcal{H}_{\rm prep}$ and $\mathcal{H}_{\rm meas}$ as their respective domains).
Thus, we must check that $\sum_{sl}\ketbra{\sigma_s(l)}=\frac{n_a}{d}\mathbbm{1}$ and 
$\sum_{tm}\ketbra{\tau_t(m)}=n_b\mathbbm{1}$, which follows from the earlier requirement
that Alice's (Bob's) multiset contain each signal (measurement) a fixed number of times. 
 
Finally, the output states can generally acquire the arbitrary phases  indicated since they will not
affect the distribution of outcomes. The phases will be important in the next section, however. The state now becomes
\begin{equation}
\frac{1}{\sqrt{n_a n_b}}\sum_{lm,st}\ket{s,l}_A\ket{t,m}_B
\left\langle\eta[{\tau_t(m)}]\big | \xi[{\sigma_s(l)}]\right\rangle e^{i[\theta(s,l)+\phi(t,m)]},
\end{equation}
where the first two systems are Alice's and the latter two Bob's, and we have used 
the notation $\ket{\xi[j]}:=\ket{\xi_j}$ and
$\ket{\eta[k]}:=\ket{\eta_k}$. Within each pair, the first system
refers to the decoding function and the second to the key 
letter. 

Now the decoding operation becomes 
trivial: simply restrict the sums over $s,t$ to only refer to proper 
function pairs. For our chosen decoding scheme, Alice and Bob 
need only exchange the results of standard basis measurements on 
$\mathcal{H}_a$ or $\mathcal{H}_b$ in order to accomplish this task. 
Averaging over all the valid function pairs is the final step, 
since Alice and Bob do not condition any of their subsequent 
actions on the particular decoding functions.

Performing this averaging procedure, one obtains the bipartite key state 
$\rho\in\mathcal{H}_{\rm key}\otimes\mathcal{H}_{\rm key}
=\mathbb{C}^r\otimes\mathbb{C}^r$. In the standard basis, 
its components are given by
\begin{equation}
\label{eq:keystate}
\rho^{\rm key}_{ij;kl}=\frac{1}{n_a n_b}\sum_{(s,t)\in T}
\Delta^{s,t}_{ij;kl}
\bracket{\eta[\tau_t(j)]}{\xi[\sigma_s(i)]}\bracket{\xi[\sigma_s(k)]}{\eta[\tau_t(l)]},
\end{equation}
where $\Delta^{s,t}_{ij;kl}=e^{i[\theta(s,i)-\theta(s,k)]}e^{i[\phi(t,j)-\phi(t,l)]}$.
Altogether, the decoding procedure defines a map from 
$\mathcal{H}_{\rm phys}\otimes\mathcal{H}_{\rm phys}$ 
to $\mathcal{H}_{\rm key}\otimes\mathcal{H}_{\rm key}$, 
whose nominal goal is to draw out the correlated portions
of the signal and measurement strings and discard the rest 
by postselection. In quantum terms, the decoding procedure
increases the entanglement of the state \emph{relative to its 
size} by simply repackaging the available entanglement
into a smaller system. (Recall that the state Alice originally 
prepared was maximally entangled, which changed 
when applying the $P$ and $M$ operations.) When the 
resulting system is highly entangled, security can be assured. 

\section{Effective Channels}
\label{sec:ec}
Now we turn to the operation of the protocol in the presence of noise. 
In principle, we must assume that all noise is due to Eve spying on the 
quantum channel. Beyond the nominal goal of concentrating entanglement, 
the decoding phase plays a pivotal role in the protocol by creating an 
\emph{effective} channel between Alice and Bob whose parameters they 
can easily estimate. Knowledge of these parameters then allows them
run the classicized CSS procedure to distill the final key.

Essentially, the averaging procedure induced by disregarding which 
key letters came from which decoding functions does all the work. 
For the moment, let us suppose that Eve tampers with each signal 
individually, performing some joint unitary operation on the signal and 
any number of ancillary systems she may care to use. For a completely
general security proof we must also consider the case in which she 
attacks blocks of signals, which we return to at the end of this section. 
The change to the signal system
itself can be described by the superoperator $\mathcal{E}=\sum_p E_p\odot E_p^\dagger$.  
Here the $E_p$ are Kraus operators or operation elements~\cite{nc00}. 
Following this channel action with the decoding map $S_AP\otimes S_BM$ yields
\begin{eqnarray}
\label{eq:keystatepresym}
\rho^{\rm key}_{ij;kl}&=&\frac{1}{n_a n_b}\sum_{p,(s,t)\in T}
\Delta^{s,t}_{ij;kl}
\bra{\eta[\tau_t(j)]}E_p\ket{\xi[\sigma_s(i)]}\nonumber\\
&&\times \bra{\xi[\sigma_s(k)]}E_p^\dagger\ket{\eta[\tau_t(l)]}.
\end{eqnarray}
[Remember that Eq.~(\ref{eq:keystate}) describes the key in the absence of channel noise.]

The symmetries of the set $T$ will now help to reduce the form of $\mathcal{E}$. 
Consider the automorphism groups $G$ for $\mathcal{S}$ and $H$ for 
$\mathcal{M}$, consisting of unitary operators $U_g$ and $V_h$ which
map the respective sets onto themselves, up to a global phase factor
for each state.  This phase factor may depend on the pair of
distillation functions $\sigma_s$ and $\tau_t$ as well.  Formally, we have
$U_g\ket{\xi_{j}}=e^{i\alpha(g,s,j)}\ket{\xi_{g(j)}}$
for some (real-valued) function $\alpha$, where $g(j)$ is used to
denote the permutation action of $G$ on $\mathcal{S}$.  Similarly for the
measurement states,
$V_h\ket{\eta_{k}}=e^{i\beta(h,t,k)}\ket{\eta_{h(k)}}$
for some function $\beta$. These operators can be applied to the
$r$-tuple specifying the destillation function $\sigma_s$, resulting in the action
\begin{equation}\label{eq:r-tup-perm}
\big(\ket{\sigma_s(0)},\dots,\ket{\sigma_s(r\minus 1)}\big)\stackrel{U_g}{\longrightarrow}
\big(U_g\ket{\sigma_s(0)},\dots,U_g\ket{\sigma_s(r\minus 1)}\big).
\end{equation}
Similarly, one obtains an action on the $r$-tuples of $\tau_t$ using $V_h$.

The symmetry group we are after is the subgroup of $G\times H$ which preserves the set $T$:
unitary operators which map pairs of $r$-tuples in $T$ to other
pairs in $T$.\footnote{Note that for $r<d$, the requirement that the right-hand side of 
Eq.~(\ref{eq:r-tup-perm}) be a valid $r$-tuple of $T$ does not fully specify $U_g$.  
So in this case, the set of operations stabilizing $T$ might be even larger.}
Call this group Aut$(T)^*$; it is a subgroup of the full automorphism group of $T$. 
In the BB84 protocol, for instance, the $r$-tuples can be transformed into
one another by a 45$^\circ$ rotation: 
\begin{equation}
\begin{array}{ccc}
(|,\textrm{---}) & \,\rightarrow\, & (\diagup,\diagdown)\\
\uparrow & & \downarrow\\
(\diagdown,\diagup) & \leftarrow & (\textrm{---},|).
\end{array}
\end{equation}
Additionally, the ordering in each pair can be separately reversed by suitable reflections.
Altogether this produces a symmetry group with eight elements.

Using Aut$(T)^*$ allows us to shift the sum over $T$ in Eq.~(\ref{eq:keystatepresym})
to a sum over the group elements, replacing each particular $(s,t)$ by 
$\big(g(x),h(y)\big)$ for some fiducial pair $(x,y)$. If the group Aut$(T)^*$ is transitive, then the entire sum can 
be rewritten in this manner. 
In case the orbit visits pairs multiple times, 
which is equivalent to the existence of a stabilizer subgroup
of Aut$(T)$ acting trivially on the fiducial pair, 
several copies of the sum are generated which can be fixed by renormalizing. 
On the other hand, if the group is not transitive, then many fiducial pairs will be 
required so that their orbits completely cover $T$. 

We are interested in the representation of Aut$(T)^*$ 
by operators of the form $U_g\otimes V_h$, which is 
is generally projective since phase factors make no difference to the quantum state. 
For the same reason, the decoding functions are also susceptible to rephasing.
To keep matters under control, we can put the latter phases to work against the former by
setting $\theta(s,j)=\alpha(g,x,j)$ for $s=g(x)$ and $\phi(t,k)=\beta(h,y,k)$ for $t=h(y)$. 
In the case that Aut$(T)^*$ is transitive, the simplified density matrix elements are
\begin{eqnarray}
\label{eq:finalkeystate}
\rho^{\rm key}_{ij;kl}&=&\frac{
\Delta^{x,y}_{ij;kl}
}{n_a n_b}\sum_{p,(g,h)\in\textrm{Aut}(T)^*}
\bra{\eta[\tau_{y}(j)]}V^\dagger_h E_pU_g\ket{\xi[\sigma_{x}(i)]}\nonumber\\
&&\times\bra{\xi[\sigma_{x}(k)]}U_g^\dagger E_p^\dagger V_h 
\ket{\eta[\tau_{y}(l)]}.
\end{eqnarray}
Thus, the symmetry of the decoding map induces an effective channel between Alice and Bob, 
described by the  symmetrized superoperator
\begin{equation}
\label{eq:supopsym}
\mathcal{E}_{\rm sym}=\sum_{p,(g,h)\in\textrm{Aut}(T)^*} V^\dagger_h E_p U_g \odot U^\dagger_g E_p^\dagger V_h.
\end{equation}
This symmetry reduces Eve's possible interference with the
effective channel. To determine the possible forms of $\mathcal{E}_{\rm sym}$, 
first express it as the output of a symmetrization 
super-superoperator $\mathcal{R}$:
\begin{equation}
\mathcal{E}_{\rm sym}=\sum_{(g,h)} 
(V^\dagger_h\odot V_h)\circ \mathcal{E}\circ(U_g \odot U^\dagger_g)=\mathcal{R}[\mathcal{E}].
\end{equation}
Appendix A details a method of using tensor products to represent superoperators by means of the 
isomorphism $A\odot B\rightarrow B^T\otimes A$, which we can use to represent Eve's effective action as  
\begin{eqnarray}
\mathcal{E}_{\rm sym}&\simeq& \sum_{p,(g,h)} V^T_h E_p^* U_g^* \otimes V^\dagger_h E_p U_g\nonumber\\
&=&\sum_{(g,h)}(V_h^*\otimes V_h)^\dagger\left(\sum_{p}E_p^*\otimes E_p\right)(U_g^*\otimes U_g)\nonumber\\
&=&\sum_{(g,h)}(V_h^*\otimes V_h)^\dagger\mathcal{E}(U_g^*\otimes U_g).
\end{eqnarray}
This reduces the symmetrization action to a superoperator itself, and we can iterate the process to write it directly as 
$\mathcal{R}\simeq\sum_{(g,h)}(U_g^\dagger\otimes U_g^T)
\otimes(V_h^T\otimes V_h^\dagger)$. $\mathcal{R}$ is Hermitian, which 
follows from the fact that the terms in the sum are group elements and 
each element is paired with its conjugate, avoiding difficulty with
the projective representation. Group composition implies that $\mathcal{R}$ 
is idempotent, up to a constant of proportionality.
Thus all possible effective channel superoperators belong
to the trivial eigenspace: $\mathcal{R}[\mathcal{E_{\rm sym}}]=\mathcal{E_{\rm sym}}$, 
a drastic reduction in the possible forms of Eve's tampering. 

The expression for the bipartite key state can be further simplified using the $\#$ operation also defined in Appendix A. Letting 
$S_A^x=\sum_{k}e^{i\theta(x,k)}\ket{k}\bra{\xi^*[\sigma_x(k)]}$ (note the conjugated state) and
$S_B^y=\sum_{k}e^{i\phi(y,k)}\ket{k}\bra{\eta[\tau_y(k)]}$, direct calculation leads to the simple expression
\begin{equation}
\label{eq:keysimple}
\rho^{\rm key}_{AB}=(S_A^x\otimes S_B^y)(\mathcal{I}\otimes\mathcal{E}_{\rm 
sym}[\ket{\Phi}_{AB}\bra{\Phi}])(S_A^x\otimes S_B^y)^\dagger,
\end{equation}
where, again, the state $\ket{\Phi}$ from Eq.~(\ref{eqn:phient}) is maximally entangled. 
Instead of averaging over the different decoding possibilities, 
now only the ``fiducial'' decoding is applied, but to the output of a suitably averaged 
channel. The end result of this analysis is to identify and delineate 
the two tasks performed by the decoding: The fiducial decoding operators 
$S_A^x$ and $S_B^y$ characterize the entanglement-enhancing 
abilities, while the effective channel operator $\mathcal{E}_{\rm sym}$
encapsulates the noise simplifications.

In case Aut$(T)^*$ is not transitive, we need only make a small
modification to the above procedure. The set $T$ is 
partitioned into disjoint orbits, and instead of choosing one fiducial decoding $(x,y)$, 
we will need one from each orbit in order to cover all of $T$. 
The final key state then contains contributions from every orbit. 
For the security analysis, we relax the condition that Alice and Bob 
throw away information regarding which decoding was used and instead treat these terms
separately. Each orbit then gives rise to an effective channel superoperator,
and we will take the worst case. 

This concludes the fully quantum-mechanical formulation of the decoding portion of the protocol.
The further steps of information reconciliation and privacy amplification can
be given a quantum formulation as a CSS-based entanglement 
distillation procedure~\cite{gp01,gl03}, which is applied to the output of the effective channel.
Distillation of maximally entangled states then assures the privacy of the key.
Given the channel parameters, the rate bounds of the CSS codes (along with
the probability of successful decoding) determine the key generation rate
of the QKD protocol. The CSS codes bring their own symmetries to the 
procedure as well, digitizing the effective channel into a Pauli channel. 

The relevant noise probabilities of the effective channel are given by overlaps with the various generalized Bell states:
\begin{equation}
b_{jk}=\bra{\beta_{jk}}\rho^{\rm key}\ket{\beta_{jk}}.
\end{equation}
For qubit-based keys, the states $\ket{\beta_{jk}}$ are the four Bell states; 
in general, they are the complete set of maximally entangled
states generated by the action of generalized Pauli operators $X^jZ^k$ 
on half the canonical maximally entangled state. Unfortunately, Alice and Bob 
do not have independent access to all these \emph{noise} probabilities. Instead, they can 
only obtain an estimate of the \emph{error} probability of the decoded keys by directly 
comparing some small fraction of them. This probability $\varepsilon$ is the 
sum of contributions from all generalized Pauli operators which are not purely of $Z$-type---i.e.,
\begin{equation}
\varepsilon=\sum_{j=0}^{d-1}\sum_{k=1}^{d-1}b_{jk}.
\end{equation}
The goal is to determine the $b_{jk}$ as functions of the error rate $\varepsilon$
or, failing that, at least find upper bounds. Then, given the Pauli channel, bounds on the rate 
of random hashing can be used to infer the secure error-rate threshold 
of the key distribution protocol~\cite{bdsw96}. 

The preceding applies when Eve performs a collective attack, interacting
with signals independently and identically. However, to establish unconditional
security of the protocol, we must consider the most general attack, called 
a coherent attack, in which Eve coherently manipulates all of the signals.
By a slight modification of the protocol we may ensure that if the protocol
is secure against collective attacks, then it is also secure against coherent 
attacks. 

The modification requires Alice and Bob to randomly reorder their signal and 
measurement data. This ensures that the error rate found by sampling 
some of the resulting key bits is representative of the error rate in the unsampled
key. This gives them direct estimates of some of the noise probabilities,
and for those which are not directly sampled, Azuma's inquality ensures that
if a relation such as $b_{jk}<f_{jk}(\varepsilon)$ holds for every key letter, then
the frequencies observed in a long sequence also obey this constraint~\cite{a67,btblr05}. 
Since the efficacy of random hashing depends on these frequencies, 
arbitrary correlations between signals pose no additional difficulties~\cite{gl03}.

One loose end remains to be tied up. In the simplified expression for the key, 
Eq.~(\ref{eq:keysimple}), some phase freedom remains in the operators
$S^x$ and $S^y$. These phases can make a difference in the secure error 
threshold of the protocol even though they have no influence on the distribution
of signal and measurement data. Though seemingly improper at
first glance, this effect is due to an inherent flexibility Alice and Bob have in 
constructing the CSS-based entanglement purification scheme.  
In canonical form, the CSS code is built from eigenspaces of products of the 
operators $X=\sum_j \ket{j+1}\bra{j}$ and $Z=\sum_k \omega^k \ketbra{k}$,
where $\omega=e^{2\pi i/d}$. However, Alice and Bob only ever actually 
measure in the standard $Z$ basis, meaning they are free to alter the $X$ 
operator in any manner consistent with the stabilizer formalism. 
In particular, they can equally well substitute $\widetilde{X}=\sum_j e^{i\psi_j}\ket{j+1}\bra{j}$ 
for $X$ without changing the crucial relationship $Z\widetilde{X}=\omega\widetilde{X}Z$. 
The altered Pauli operators give rise to a rephased variant of the maximally entangled
states,
\begin{equation}
\ket{\widetilde{\beta}_{jk}}=\mathbbm{1}\otimes \widetilde{X}^jZ^k\ket{\Phi}=
\frac{1}{\sqrt{d}}\sum_{l}\omega^l\exp\left[i\sum_{m=0}^{j-1}\psi_{l+m}\right]\ket{l}\ket{l+j}.
\end{equation}
For instance, in the case of two two-level systems, the general set 
of maximally entangled states reads $\ket{00}\pm\ket{11}, 
e^{i\psi_0}\ket{01}\pm e^{i\psi_1}\ket{10}$. Thus, 
altering the phases appearing in $S^x$ and $S^y$ 
can indeed change the distribution of noise $b_{jk}$ without 
affecting the distribution of signals and measurements $p_{jk}$. 

\section{Security of Spherical Code QKD Protocols in Small Dimension}
\label{sec:secproofs}

The preceding gives a general method for establishing the unconditional 
security of protocols exhibiting a high degree of symmetry. One needs
(merely) to find the relevant automorphism groups and then straightforwardly 
compute the $b_{jk}$ distribution to determine the secure error threshold for 
any given protocol. To demonstrate this technique, we turn our attention 
to quantum key distribution protocols employing equiangular spherical 
code signal states in low dimensions. 

\subsection{Qubit tetrahedron}
In the tetrahedral protocol of~\cite{renes04}, Alice's signal 
qubits are given by four states whose Bloch vectors form a regular 
tetrahedron. Bob's measurement states correspond to the inverse 
(in the sense of the Bloch sphere) of this tetrahedron, so that each of 
his outcomes rules out one potential signal. Alice decodes two of the
states into a logical bit and announces which ones to Bob. If his 
measurement rules out one of the possibilities, he can determine the bit;
the whole procedure succeeds with probability one-third in the absence
of noise. Alice's decoding functions are equivalent to ordered pairs
of signal states, of which there are 12. Since Bob's successful
decoding function is completely specified by Alice's, 12 possible
decoding combinations exist in total and only one-way communication
from Alice to Bob is required. The automorphism group corresponds to
$A_4$, the alternating group on four elements, and can be projectively
represented on $\mathbb{C}^2$ for both parties using the following two
generators:
\begin{equation}
\sigma_x=\left(\begin{array}{cc}0&1\\1&0\end{array}\right),\quad \frac{1}{\sqrt{2}}\left(\begin{array}{cc}1 & -i\\1 &i\end{array}\right).
\end{equation}
This group representation corresponds to using the tetrahedron generated from the fiducial state
$(\sqrt{3+\sqrt{3}}\ket{0}+\sqrt{3-\sqrt{3}}e^{i\pi/4}\ket{1})/\sqrt{6}$ for the signals.  

From the automorphism group it is easy to calculate the trivial eigenspace of 
$\mathcal{R}$, which is in this case spanned by only two superoperators.
By appropriate use of the $\#$ operation, the output of the channel for 
maximally entangled input can be expressed as a linear combination
of the identity operator and the maximally entangled state again---i.e., the 
depolarizing channel. Next, a fiducial decoding consisting 
of restriction to the fiducial signal state and its image under the first $A_4$ generator $\sigma_x$ can
be applied and the error probabilities of the resulting state can be tabulated. 
In terms of the depolarization rate $p$, they are given by 
$b_{01}=b_{11}=2b_{10}=p/(2+p)$ and $b_{00}=1-b_{01}-b_{10}-b_{11}$, where all
phases appearing in the fiducial decoding operators were set to zero. 
Finally, we can apply the random hashing bound on the number of 
distillable maximally entangled states $S(\rho)$ from an input state
$\rho$ with diagonal entries in the Bell basis given by $b_{jk}$: 
$S(\rho)\geq 1-H(\{b_{jk}\})$, where $H$ is the Shannon entropy. 
From this one obtains a threshold depolarization rate of approximately 16.69\%, 
corresponding to an error rate $\varepsilon=3p/(4+2p)$ of approximately 11.56\%. 

\subsection{Qutrit spherical code protocols}
For qutrits---three-level quantum systems---there are four possible 
equiangular spherical code signal ensembles Alice could choose from, with $n=$4, 6, 7, and 9
elements, respectively. A myriad of protocols exist using these as signals, 
but here we confine our attention to those for which Bob's measurement outcomes
are orthogonal to two signal states and the goal of the decoding step 
is to establish one bit. The latter requirement means that the decoding functions have
support on only two signals at a time, or in other words, Alice informs Bob that the signal sent is one of only two possibilities.
The set $T$ consists of function pairs corresponding to the cases in which Bob's measurement outcome and Alice's 
announcement allow him to determine which signal she sent. Since the raw key alphabet consists of just two letters, 
Alice and Bob can make use of \emph{qubit} CSS codes to perform error-correction and privacy amplification.

The technique of having Bob's measurement repudiate some of Alice's signals was introduced briefly in~\cite{renes05}. 
Letting $\Pi_j=\ketbra{\xi_j}$, we can formulate the measurement as 
\begin{equation}
\ketbra{\eta_{j,k}}\propto \mathbbm{1}-\frac{\Pi_j+\Pi_k-\{\Pi_j,\Pi_k\}}{1-{\rm Tr}[\Pi_j\Pi_k]}.
\end{equation}
Since for spherical codes the denominator does not depend on $j$ and $k$, 
the set of projectors can easily be found to sum to the identity operator.

For each protocol we attempt to find Aut$(T)^*$ and from this extract
the possible outputs of the corresponding effective channel for each orbit in $T$; 
Aut$(T)^*$ is nontransitive for all these protocols.
Then the phases of the canonical decoding operators 
must be judiciously chosen to find the best secure error threshold. 
In the first three protocols $n=4,6,7$ it is necessary to 
give up on random hashing directly and retreat to finding a 
CSS code which can correct the bit and phase errors independently, 
for there are too many parameters to determine the relationships 
between the various Pauli errors exactly. 
This strategy was also used in the security analysis of the trine protocol~\cite{btblr05}.
When $n=9$ the effective channel is again a depolarizing channel and therefore 
the better technique of random hashing can be used. 

\begin{table*}[ht]
\begin{tabular}{cp{.5in}cp{.5in}c}
Spherical code protocol && Threshold error rate && Sufficient threshold fidelity\\\hline
\rule[-3mm]{0mm}{8mm}$\phantom{.}$[4,2,2,1] && 11.56\% && 0.917\\
$\phantom{.}$[4,3,2,2] && $\phantom{0}$8.90\% && 0.881\\
$\phantom{.}$[6,3,2,2] && 11.00\% && 0.844\\
$\phantom{.}$[7,3,2,2] && 10.37\% && 0.916\\
$\phantom{.}$[9,3,2,2] && 11.80\% && 0.843
\end{tabular}
\caption{Threshold error rates and fidelities for the qubit tetrahedron protocol and the 
four spherical-code-based key distribution protocols in three dimensions. The protocols are
named according to the convention $[n,d,k,m]$ where $n$ is the number 
of signals, $d$ is the dimension of the associated vector space,
$k$ is the number of possible remaining signals after Alice announces 
the decoding information (i.e.,$k$ is the size of the key alphabet), and $m$ is the number
of signals which are ruled out by Bob's measurement. The threshold 
error rate is the maximum secure error rate of the key, while the sufficient
threshold fidelity is a upper bound on the corresponding fidelity of the 
output state of the symmetrized channel with the maximally entangled input.
Fidelities beyond this limit are sufficient for key creation. 
}
\label{tab:tritqkd}
\end{table*}

Table~\ref{tab:tritqkd} shows the threshold error rates 
and sufficient threshold fidelities for these protocols, 
whose details are laid out in Appendix B.
One might expect a trend to higher tolerable error rates 
and minimal fidelities with increasing number of signals, 
but the seven-element spherical code breaks rank, requiring 
the cleanest channel. A quick check of the appendix reveals 
the reason: there are 1050 possible decoding combinations, 
but the largest known automorphism group has only 42 elements, 
yielding 25 distinct orbits. Thus, a fairly large
mismatch exists between the symmetry of the decoding 
combinations and those realized by action on the signals, 
which simply does not restrict the channel as much as in the 
other cases. However, it should be noted that the full automorphism 
group might be larger. 

\section{Discussion and Conclusion}
\label{sec:dc}
By treating the decoding step of prepare and measure protocols quantum mechanically, 
we have presented a general formalization which enables the 
application of the entanglement-distillation proof
technique to a broad class of key distribution protocols. 
Additionally, we gain insight into the general role decoding plays in
quantum key distribution. On the one hand, the symmetries 
of the signal and measurement states are found to play a clear and direct role
in security proofs. Applying this machinery to the 
symmetries of various equiangular spherical codes in 
two and three dimensions yielded the secure error threshold 
for the associated protocols. On the other hand, 
treating the decoding step in more detail reveals the 
mechanism by which Alice and Bob are able to estimate 
the various noise parameters: The decoding step creates 
an ``effective'' quantum channel between Alice and Bob, whose
noise properties they can more easily estimate than those of the physical channel. 
The decoding creates a sort of logical communication 
layer embedded within the physical layer of actual signals. 
With this formulation in place, we can now begin to consider 
more complicated decoding strategies: 
two-way communication, block-wise and set-wise decoding, 
as well as more general ``imperfect'' protocols whose 
decoding schemes which produce nonmaximally entangled states. 

\section*{ACKNOWLEDGEMENTS}
The authors acknowledge helpful conversations with 
Bryan Eastin, Marcos Curty, and Norbert L\"utkenhaus. 
This work was supported by BMBF Project No.~01/BB01B, 
the SECOQC project of the European Commission, and
DFG Schwerpunktprogramm Quanten-Informationsverarbeitung 
(SPP 1078), Projekt AQUA (Be~887/13).

\appendix
\section{Superoperator representations}
In expressing superoperators, one convention 
is to use ``$\odot$'' as a placeholder for the input 
operator---i.e., $(A\odot C)[B]=ABC$. However, this 
makes representations cumbersome. Instead one 
would like to express the superoperator as a 
matrix and the input operator as a vector. This is 
easily done, albeit in two distinct ways. First, note that 
we can flatten the operator $B$ into a vector by applying it to 
half of a maximally entangled state, like so: $B\rightarrow (\mathbbm{1}\otimes B)\ket{\Phi}$. 
This action is called the VEC map. The action of $(A\odot C)$ on $B$ 
then becomes multiplication of VEC$(B)$ by the operator $C^T\otimes A$. 
Another matrix representation of superoperators can be obtained by 
applying the superoperator to half of a maximally entangled state. This furnishes
a representation similar to the way VEC produces a vector from an 
operator, so the map is termed OP. 
Here one finds immediately that 
$\big(\mathcal{I}\otimes (A\odot C)[\Phi]\big)_{ij;kl}=a_{ji}c_{kl}$, which is 
related to the $C^T\otimes A$ representation by simply 
interchanging the first and last indices $i$ and $l$. 
This ``partial transposition'' can be denoted by $\#$, as in
\begin{equation}
\label{eq:sharpdef}
\left[{\rm OP}(A\odot C)\right]_{ij;kl}=\left[(C^T\otimes A)^\#\right]_{ij;kl}.
\end{equation}
A more detailed account of this superoperator sleight of hand can be found in~\cite{caves99}.

\section{Qutrit Security Details}
In this appendix we list the ingredients which are required to 
complete the security proof for the qutrit spherical code protocols, including
the signal states, the automorphism group, and the phases of the decoding operators. 
\subsection{[4,3,2,2]}
Beginning with the vector $\frac{1}{\sqrt{3}}(\ket{0}+\ket{1}+\ket{2})$, 
four equiangular states labeled $0,1,2,3$
can be generated from it by application
of the operators
\begin{equation}
\left(\begin{array}{ccc}1&0&0\\0&0&-1\\0&-1&0\end{array}\right),\quad 
\left(\begin{array}{ccc}0&0&-1\\0&1&0\\-1&0&0\end{array}\right),\quad 
\left(\begin{array}{ccc}0&-1&0\\-1&0&0\\0&0&1\end{array}\right),
\end{equation}
in this order. These operators also generate the automorphism group, 
which is a representation of $S_4$. Take the fiducial decoding 
$(0,1)$ for Alice and $(\{1,2\},\{0,3\})$ for Bob, where the 
two sets identify the states ruled out by his measurement.
Then Alice's phases can both be set to zero, while Bob 
must choose them to have a difference of $\pi$. From this it 
follows that $e_{\rm phase}\leq \frac{3}{2} e_{\rm bit}$, 
which then leads to 8.90\% by the CSS rate bound
$1-h_2(e_{\rm bit})-h_2(e_{\rm phase})$. Here $h_2$ is the binary Shannon entropy.
This also holds when we consider three other decodings for Bob: 
$(\{1,3\},\{0,2\})$, $(\{1,2\},\{0,2\})$, $(\{1,3\},\{0,3\})$. Under
the group action these four fiducial decodings cover all
of the 48 possibilities.

\subsection{[6,3,2,2]}
First let $\phi$ be the golden ratio $(1+\sqrt{5})/2$. Then starting with
$(\sqrt{\phi}\ket{1}+\sqrt{\phi\!-\!1}\ket{2})/5^{1/4}$, six equiangular 
vectors are generated by the group formed from the first two of the following:
\begin{equation}
\left(\begin{array}{ccc}0&0&1\\1&0&0\\0&1&0\end{array}\right), \quad
\left(\begin{array}{ccc}1&0&0\\0&1&0\\0&0&-1\end{array}\right),\quad 
\frac{1}{2}\left(\begin{array}{ccc}\phi&1-\phi&1\\1-\phi&1&\phi\\1&\phi&1-\phi\end{array}\right).
\end{equation}
Adding the final operator gives the full automorphism group, 
which is a projective representation of $A_5$. One can then derive that $e_{\rm phase}\leq e_{\rm bit}$. 
\subsection{[7,3,2,2]}
Letting $\eta=\exp[2\pi i/7]$, the seven-element spherical code in three dimensions can be generated by repeated application 
of the first of the operators
\begin{equation}
\left(\begin{array}{ccc}\eta&0&0\\0&\eta^2&0\\0&0&\eta^4\end{array}\right),\quad
\left(\begin{array}{ccc}0&1&0\\0&0&1\\1&0&0\end{array}\right)\quad
\end{equation}
to the vector $\frac{1}{\sqrt{3}}(\ket{0}+\ket{1}+\ket{2})$. 
The second stabilizes the starting vector, along with the antiunitary operation of complex
conjugation in the standard basis. Altogether this yields a group of order 42, using which one derives that 
$e_{\rm phase}<\frac{9}{8} e_{\rm bit}$. 
\subsection{[9,3,2,2]}
Let $\omega=e^{2\pi i/3}$. Forming the nine element group from the first two of the generators
\begin{eqnarray}
&&\left(\begin{array}{ccc}0&0&1\\1&0&0\\0&1&0\end{array}\right),\quad
\left(\begin{array}{ccc}1&0&0\\0&\omega&0\\0&0&\omega^2\end{array}\right),\nonumber\\
&&\left(\begin{array}{ccc}1&0&0\\0&\omega^2&0\\0&0&\omega^2\end{array}\right),\quad 
\frac{1}{\sqrt{3}}\left(\begin{array}{ccc}1&\omega^2&\omega^2\\1&1&\omega\\1&\omega&1\end{array}\right),
\end{eqnarray} 
and applying them to the fiducial vector $(\ket{1}-\ket{2})/\sqrt{2}$ generates the SICPOVM. The latter two generators
stabilize the fiducial vector and enlarge the automorphism 
group to consist of 216 elements which are isomorphic to the 
Shephard-Todd reflection group number 25 modulo its center~\cite{ShTo54}. 
This group gives rise to a depolarizing effective channel, and the various errors are found 
to obey the relations $b_{1,0}=2 b_{0,1}(6-\sqrt{3})/15$ and $b_{1,1}=2b_{0,1}(6+\sqrt{3})/15$.


\begin{thebibliography}{99}
\bibitem{lc99} H.-K. Lo and H.~F. Chau, Science {\bf 283}, 2050 (1999). 

\bibitem{sp00} P.~W. Shor and J. Preskill, Phys. Rev. Lett. {\bf 85}, 441 (2000).

\bibitem{css} A.~R. Calderbank and P.~W. Shor, Phys.~Rev.~A {\bf 54}, 1098 (1996); 
A.~M. Steane, Proc. R. Soc. London, Ser. A {\bf 452}, 2551 (1996).

\bibitem{bb84} C.~H. Bennett and G. Brassard, 
in {\it Proceeding of the IEEE International Conference on Computers, Systems, and Signal Processing, Bangalore, India, 1984},  (IEEE, New York, 1984), pp. 175--179.

\bibitem{bruss98} D.~Bru\ss, Phys.~Rev.~Lett.~{\bf 81}, 3018 (1998).

\bibitem{l01} H.-K. Lo, Quantum Inf.~Comput.~{\bf 1}, 81 (2001).

\bibitem{Filtering} N.~Gisin, Phys.~Lett.~A {\bf 210}, 151 (1996); 
M. Horodecki, P. Horodecki, and R. Horodecki, Phys. Rev. Lett. {\bf 78}, 574 (1997).

\bibitem{b92} C.~H. Bennett, Phys. Rev. Lett. {\bf 68}, 3121 (1992).

\bibitem{tki03}  K. Tamaki, M. Koashi, and N. Imoto, Phys. Rev. Lett. {\bf 90}, 167904 (2003).

\bibitem{pbc00} S. Phoenix, S. Barnett, and A. Chefles, J. Mod. Opt. {\bf 47}, 507 (2000).

\bibitem{renes04} J.~M. Renes, Phys. Rev. A {\bf 70}, 052314 (2004).

\bibitem{btblr05} J.~C.~Boileau, K.~Tamaki, J.~Batuwantudawe, R.~Laflamme, and J.~M.~Renes, Phys. Rev. Lett. {\bf 94}, 040503 (2005).

\bibitem{maurer93}U.~M.~Maurer, IEEE Trans. Inf. Theory {\bf 39}, 733 (1993).

\bibitem{nc00} M.~A.~Nielsen and I.~L.~Chaung, \emph{Quantum Computation and Quantum
Information} (Cambridge University Press, Cambridge, UK, 2000).

\bibitem{gp01} D.~Gottesman and J.~Preskill, Phys.~Rev.~A {\bf 63}, 022309 (2001).

\bibitem{gl03} D. Gottesman and H.-K. Lo, IEEE Trans. Inf. Theory {\bf 49}, 457 (2003).

\bibitem{bdsw96} C.~H. Bennett, D.~P. DiVincenzo, J.~A. Smolin, and W.~K. Wootters, Phys. Rev. A {\bf 54}, 3824 (1996).

\bibitem{a67} K. Azuma, Tohuku Math. J. {\bf 19}, 357 (1967).

\bibitem{renes05} J.~M.~Renes, Quantum Inf. Comput. {\bf 5}, 81 (2005).

\bibitem{caves99} C.~M.~Caves, J. Supercond. {\bf 12}, 707 (1999).


\bibitem{ShTo54} G.~C.~Shephard and J.~A.~Todd,  
Canadian J. Math. {\bf 6} 274 (1954). 

\end{thebibliography}
\end{document}